%% file: kramer.tex
\documentclass[vecphys]{svmult}

\usepackage{makeidx}         
\usepackage{graphicx}        
\usepackage{multicol}        
\usepackage[bottom]{footmisc}

\makeindex             

\newcommand{\msun}{M$_{\odot}$}   
   
\begin{document}

\title*{A complete CO 2-1 map of M51 with HERA}
\author{
  Carsten\,Kramer\inst{1} \and 
  Marc\,Hitschfeld\inst{1} \and 
  Karl\,F.\,Schuster\inst{2} \and 
  Santiago\,Garcia-Burillo\inst{3} \and 
  Bhaswati\,Mookerjea\inst{4}
}
\authorrunning{Kramer, Hitschfeld et al.}
\institute{
  KOSMA, I. Physikalisches Institut, Universit\"at zu K\"oln,   
  Z\"ulpicher Stra\ss{}e 77, 50937 K\"oln, Germany
  \texttt{kramer@ph1.uni-koeln.de}
\and 
  IRAM, 300 Rue de la Piscine, F-38406 S$^t$ Martin d'H\`{e}res, France 
  \texttt{schuster@iram.fr}
\and
  Centro Astronomico de Yebes, IGN, E-19080 Guadalajara, Spain
  \texttt{burillo@oan.es}
\and 
  Department of Astronomy, University of Maryland, College Park, MD 20742, USA 
  \texttt{bhaswati@astro.umd.edu}
}
%
%
\maketitle


\begin{abstract}
  The nearby, almost face-on, and interacting galaxy M51 offers an
  excellent opportunity to study the distribution of molecular gas and
  the mechanisms governing the star formation rate.  We have created a
  complete map (Fig.\,\ref{fig_m51}) of M51 in $^{12}$CO 2--1 at a
  resolution of $11''$ corresponding to 450\,kpc using HERA at the
  IRAM-30m telescope.  In \cite{schuster2006} we have combined these
  data with maps of HI and the radio-continuum to study the star
  formation efficiency, the local Schmidt law, and Toomre stability of
  the disk in radial averages out to radii of 12\,kpc. Here, we also
  discuss the distribution of giant molecular associations and its
  mass spectrum, in comparison with similar studies in the literature.
\end{abstract}

\section{CO, HI, and the radio continuum}

\begin{figure}
\centering
\includegraphics[angle=-90,width=12cm]{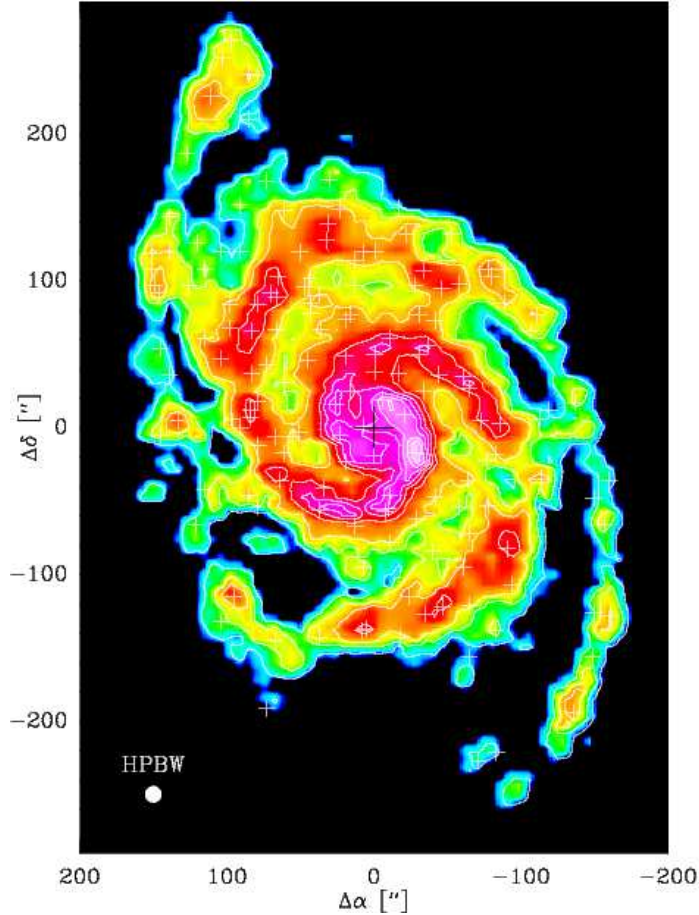}
\caption{
  Map of $^{12}$CO 2--1 integrated intensities in Kkms$^{-1}$ showing
  M51, i.e. NGC\,5194 and its companion galaxy NGC\,5195 in the
  north-east. Crosses mark the center positions of the identified 155
  giant molecular associations (GMAs).  }
\label{fig_m51}       
\end{figure}

\begin{figure}
\centering
\includegraphics[angle=-90,width=10cm]{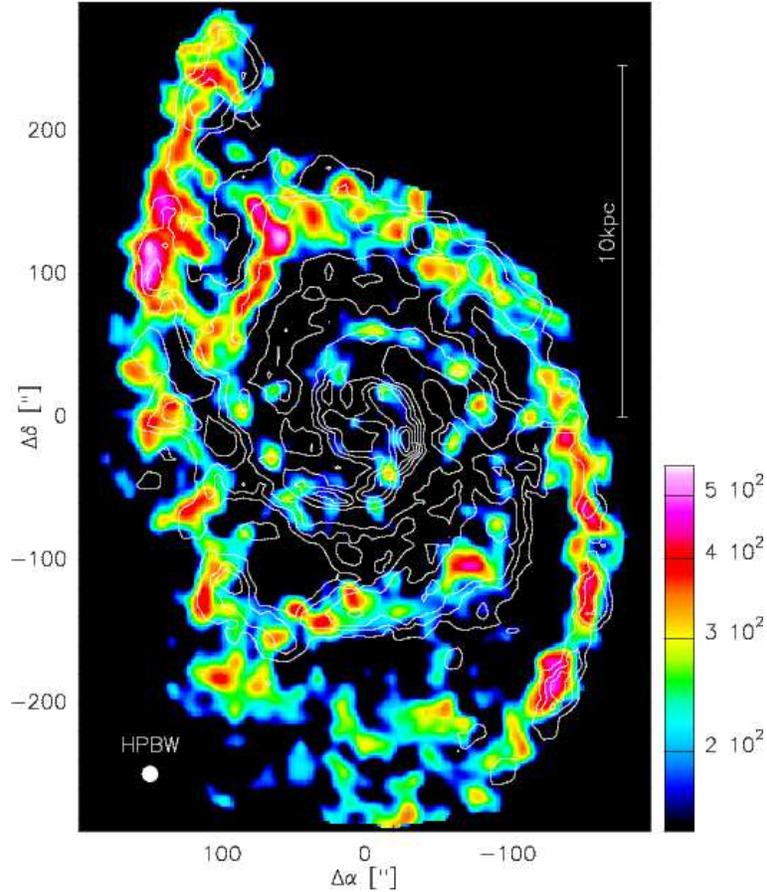}
\caption{
  VLA map of integrated HI intensities [Jy/beam] at $13''$ resolution 
  \cite{rots1990} in colors.  Contours show integrated $^{12}$CO 2--1
  intensities (cf.\,Fig.\,\ref{fig_m51}).
}
\label{fig_m51hi}       
\end{figure}

\begin{figure}[h]   
\centering   
\includegraphics[angle=-90,width=7.5cm]{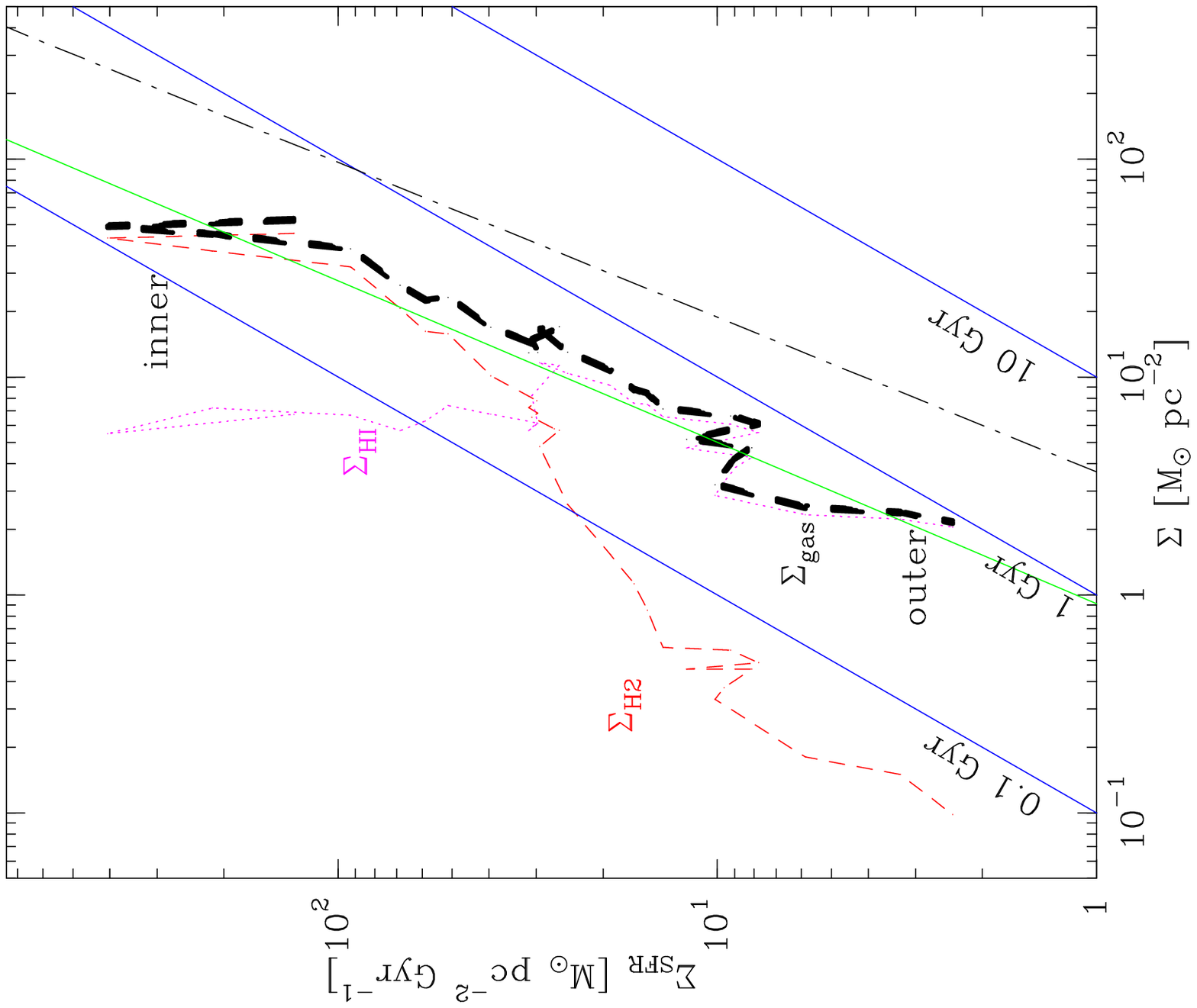}
\caption{Radially averaged star formation   
  rate per unit area, $\sum_{\rm SFR}$, versus surface densities of
  the total gas $\sum_{\rm gas}$, and of H$_2$ and HI only.  The
  solid green line is the local Schmidt-law found in M51. The
  dashed-dotted black line is the global Schmidt-law found by
  \cite{kennicutt1998}.  Drawn blue lines represent lines of constant
  gas depletion time or star formation efficiency. }
\label{fig_schmidt}   
\end{figure}   

M51 is an interacting, grand-design spiral galaxy at a distance of
only 8.4\,Mpc seen nearly face-on.  The emission detected with the 30m
telescope (Fig.\,\ref{fig_m51}) traces the well known two-armed spiral
pattern out to the companion galaxy NGC\,5195, which shows up brightly
in the north-east at $\sim10.5$\,kpc radial distance, and out to the
south-western tip of the second arm at the opposite side of M51.  The
outer parts of the two arms in the west and in the east appear more
fragmented than the inner parts.  The western arm especially is almost
unresolved.  Inter-arm emission is detected above the $3\sigma$ level
out to radii of about 6\,kpc.  Several spoke-like structures connect
the spiral arms radially.
The large-scale distribution of the 21\,cm line of atomic hydrogen in
M51 was analyzed by \cite{rots1990} using the VLA.  The HI emission at
$13''$ resolution (Fig.\,\ref{fig_m51hi}) is weak in the inner region
while the outer CO arms are clearly delineated in HI. These data have
been combined with a new large scale 20cm map of \cite{patrikeev2006}
which provides an extinction-free estimate of the star formation rate.

We derive a global star formation rate of 2.56\,\msun\,yr$^{-1}$.  The
total gas surface density $\sum_{\rm gas}=1.36(\sum_{{\rm
    H}_2}+\sum_{\rm HI})$ drops by a factor of $\sim20$ from
70\,\msun\,pc$^{-2}$ at the center to 3\,\msun\,pc$^{-2}$ in the
outskirts at radii of 12\,kpc.  The ratio of HI over H$_2$ surface
densities, $\sum_{\rm HI}/\sum_{\rm H2}$, increases from $\sim0.1$
near the center to $\sim20$ in the outskirts without following a
simple power-law.  The star formation rate per unit area drops from
$\sim400$\,\msun\,pc$^{-2}$\,Gyr$^{-1}$ in the starburst center to
$\sim2$\,\msun\,pc$^{-2}$\,Gyr$^{-1}$ in the outskirts. $\sum_{\rm
  gas}$ and $\sum_{\rm SFR}$ are well characterized by a local
Schmidt law $\sum_{\rm SFR}\propto\sum_{\rm gas}^n$ with a
power-law index of $n=1.4\pm0.6$ (Fig.\,\ref{fig_schmidt}).

The critical gas velocity dispersions needed to stabilize the gas
against gravitational collapse in the differentially rotating disk of
M51 using the Toomre criterion, vary with radius between 1.7 and
6.8\,kms$^{-1}$. Observed radially averaged dispersions derived from
the CO data vary between 28\,kms$^{-1}$ in the center and
$\sim$8\,kms$^{-1}$ at radii of 7 to 9\,kpc. They thus exceed the
critical dispersions by factors $Q_{\rm gas}$ of 1 to 5. Taking into
account, in addition, the gravitational potential of stars, the disk may
be critically stable.

\section{Distribution of molecular gas}

Molecular clouds are the sites of all star formation in the Milky Way
and also in external galaxies. It is therefore of great interest to
study the global distribution of star formation in entire galaxies by
studying the distribution of the molecular gas.
Nearby face-on galaxies like M51 offer the possibility to study the
distribution of molecular gas over their entire surface at high
spatial resolution and without the distance ambiguities as encountered
for the Milky Way.
In M51, the total mass of molecular material derived from the
integrated $^{12}$CO 2--1 intensities is $2\,10^9$\,\msun.  The
$3\sigma$ limit with resolutions of $11''$ and 5\,kms$^{-1}$
corresponds to a mass of $1.7\,10^5$\,\msun. 
The spatial resolution of 450\,pc resolves structures larger than
typical GMCs which we label giant molecular associations. These may be
bound clusters of GMCs as suggested by \cite{rand_kulkarni1990} or random
superpositions of GMCs \cite{gb1993no2}.
%
%
In a first attempt to study the properties of these GMAs, we have
decomposed the CO 2--1 data set into Gaussian shaped clouds using the
{\it gaussclumps} algorithm \cite{stutzki1990}. This algorithm has
been developed to study the statistical properties of the structure of
molecular clouds as seen in spectral line emission cubes and has been
applied to many Galactic clouds, e.g.  \cite{kramer1998,simonr2001}.
It iteratively fits a local Gaussian to the global maximum of the data
and subtracts it.  The algorithm fits the intrinsic, deconvolved
widths to the observed data taking into account the given angular and
spectral resolutions. It outputs the clump center positions and widths
in the two spatial and in the velocity coordinates.  It also outputs
the clump peak temperatures, as also their orientations. In the
following, we will use the term clouds or GMAs rather than clumps.

\begin{figure}
\centering
\includegraphics[angle=-90,width=6cm]{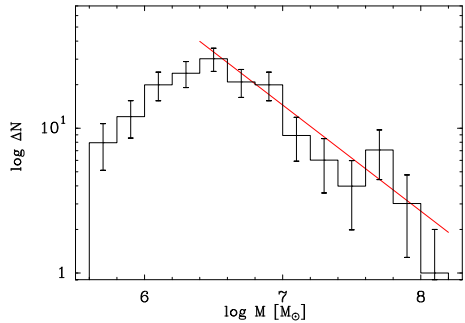}
\caption{The mass spectrum of GMAs identified in M51 shows the 
  number of sources per logarithmic mass interval.  Errorbars represent
  $\sqrt{\Delta N}$ statistical errors.  The best fitting power law of
  the form $dN/dM \propto M^{-\alpha}$ has a slope of
  $\alpha=2.0\pm0.2$.}
\label{fig_massspectrum}       
\end{figure}

\begin{table}
\centering
\caption{Mass spectra of GMCs and GMAs in external galaxies. The number of identified clouds
is given in column 4. $M_{\rm min}$ and $M_{\rm max}$ are the minimum and maximum
cloud mass detected. $M_{\rm turn}$ is the mass at the turnover of the spectrum. }
\label{tab_massspectra}       
\begin{tabular}{lrrccrrrr}
\hline\noalign{\smallskip}
Source        & Distance  & Resol. & \hspace*{0.1cm} No. \hspace*{0.1cm}  
                                     & $\alpha$      & $M_{\rm min}$ 
                                                        & $M_{\rm turn}$ 
                                                                    & $M_{\rm max}$ 
                                                                                & \hspace*{0.1cm} Lit. \\
              &           & [pc$^2$] &   & [M$_\odot$] & [M$_\odot$] & [M$_\odot$] \\
\noalign{\smallskip}\hline\noalign{\smallskip}
NGC\,4038/39  & 19\,Mpc   & \hspace*{0.1cm}
                            $310\times480$ & $100$ & $1.4\pm0.1$   & $2\,10^6$ & $5\,10^6$ & $9\,10^8$ & \cite{wilson2003} \\ 
M51           &  8.4\,Mpc & 450 & 155 & $2.0\pm0.2$   & $5\,10^5$ & $3\,10^6$ & $1\,10^8$ & \cite{hitschfeld2007} \\ 
M33           & 850\,kpc  & 50 & 148 & $2.6\pm0.3$   & $3\,10^4$ & $2\,10^5$ & $7\,10^5$ & \cite{engargiola2003} \\ 
M31           & 780\,kpc  & 90 & 389 & $1.6\pm0.2$   & $2\,10^4$ & $1\,10^5$ & $5\,10^5$ & \cite{muller2006} \\ 
LMC           & 54\,kpc   & 41 & 168 & $1.9\pm0.1$   & $4\,10^4$ & $8\,10^4$ & $3\,10^6$ & \cite{fukui2001} \\ 
%
\noalign{\smallskip}\hline
\end{tabular}
\end{table}

In M51, the algorithm decomposes 78\% of the total mass into 155
clouds. Their deconvolved, i.e. intrinsic, sizes and FWHMs are larger
than 20\% of the resolutions.  The 16 most massive GMAs with masses
between $2\,10^7$\,M$_\odot$ and $1\,10^8$\,M$_\odot$ follow the two
inner logarithmic spiral arms like beads on a string
(Fig.\,\ref{fig_m51}).  These masses exceed the masses derived by
\cite{rand_kulkarni1990} using the three-element Caltech Millimeter
Array by a factor of $\sim2$, presumably due to the slightly larger
beam size of the present study and missing short spacing information.
Such very massive molecular cloud complexes have also been found in
other galaxies. See the compilation by \cite{wilson2003} (their Table
3).
In our study of M51, GMAs are also identified along the two outer arms
extending towards the companions galaxy in the north and to the south.
At least five spoke like features connecting spiral arms are
decomposed into clouds.  Only few clouds are seen in the interarm
medium. The least massive clouds have a mass of $5\,10^5$\,M$_\odot$.

Figure\,\ref{fig_massspectrum} shows the mass spectrum of all 155
clouds. It follows a linear slope of $\alpha=2.0\pm0.2$ above the
turnover mass of $5\,10^6$\,M$_\odot$. The turnover is presumably not
intrinsic to the GMA distribution but caused by the detection limits,
similar as in Galactic clouds \cite{kramer1998}.
A slope of 2.0 agrees well, within the error, with more recent
large-scale mapping studies of Galactic clouds by e.g.
\cite{heithausen1998,kramer1998,simonr2001,heyer2001}. These studies
find a rather constant slope of $\alpha=1.6-1.9$ for cloud and clump
masses between more than $10^4$\,M$_\odot$ and $10^{-4}$\,M$_\odot$.
Todate, only few studies exist which derive mass spectra of the entire
cloud population of external galaxies. In Table \ref{tab_massspectra},
we compare the mass spectrum of M51, with those found in NGC\,4038/39,
M33, M31, and the LMC. The slope which we find in M51 is similar to
the slope found in the Andromeda Galaxy using the same algorithm,
though the study of M31 covers a lower mass range due to its
proximity. 
The slope is also similar to the slope found in the nearby dwarf
galaxy LMC. The Antennae exhibit a rather flat mass distribution with
very massive GMAs while M33 shows a very steep spectrum with
$\alpha=2.6$ similar to the initial mass function.
  

%
\input{kramer_referenc}


\printindex
\end{document}

%% file: kramer_referenc.tex
%
%

%
%


%% file: kramer.bbl
\begin{thebibliography}{99.}
%
%
%


\bibitem{schuster2006} K. Schuster, C. Kramer, M. Hitschfeld, S. Garcia-Burillo,
B. Mookerjea: accepted for publication in A\&A (2006)

\bibitem{rots1990} A.H. Rots, P.C. Crane, A. Bosma, E. Athanassoula,
  J.M. van der Hulst: ApJ, \textbf{100}, 387 (1990)

\bibitem{patrikeev2006} I. Patrikeev, A. Fletcher, R. Stepanov, R.
  Beck, E.M. Berkhuijsen, P. Frick, C. Horellou: A\&A in press (2006)

\bibitem{kennicutt1998} R.C. Kennicutt: ApJ, \textbf{498}, 541 (1998)
  
\bibitem{rand_kulkarni1990} R.J. Rand, S.R. Kulkarni: ApJL,
  \textbf{349}, L43 (1990)
  
\bibitem{gb1993no2} S. Garcia-Burillo, F. Combes, M. Gerin: A\&A,
  \textbf{274}, 148 (1993)
  
\bibitem{stutzki1990} J. Stutzki, R. Guesten: ApJ, \textbf{356}, 513
  (1990)

\bibitem{kramer1998} C. Kramer, J. Stutzki, R. Roehrig, U. Corneliussen:
  A\&A, \textbf{329}, 249 (1998)
  
\bibitem{wilson2003} C.D. Wilson, N. Scoville, S.C. Madden, V.
  Charmandaris: ApJ, \textbf{599}, 1049 (2003)
  
\bibitem{hitschfeld2007} M. Hitschfeld, C. Kramer, K. Schuster, S.
  Garcia-Burillo, J. Stutzki, B. Mookerjea: A\&A, in prep. (2007)
  
\bibitem{engargiola2003} G. Engargiola, R.L. Plambeck, E. Rosolowsky,
  L.  Blitz: ApJS, \textbf{149}, 343 (2003)
  
\bibitem{muller2006} S. Muller: ``Molecular gas in the Andromeda
  Galaxy: properties of the molecular clouds'', this conference

  
\bibitem{fukui2001} Y. Fukui, N. Mizuno, R. Yamaguchi, A. Mizuno, T.
  Onishi: PASP, \textbf{53}, 41 (2001)

  
\bibitem{heithausen1998} A. Heithausen, F. Bensch, J. Stutzki, E.
  Falgarone, J.F. Panis: A\&A, \textbf{331}, 65 (1998)
  
\bibitem{simonr2001} R. Simon, J.M. Jackson, D.P. Clemens, T.M. Bania,
  M.H. Heyer: ApJ, \textbf{551}, 747 (2001)
  
\bibitem{heyer2001} M.H. Heyer, J.M. Carpenter, R.L. Snell: ApJ,
  \textbf{551}, 851 (2001)




  

\end{thebibliography}
